\documentclass[sigconf,nonacm]{acmart}
\AtBeginDocument{%
  }

\setcopyright{acmlicensed}

\copyrightyear{2026}
\acmYear{2026}
\setcopyright{cc}
\setcctype{by}
\acmConference[MM '26] {Proceedings of the 34th ACM International Conference on Multimedia}{November 10--14, 2026}{Rio de Janeiro, Brazil.}
\acmBooktitle{Proceedings of the 34th ACM International Conference on Multimedia (MM '26), November 10--14, 2026, Rio de Janeiro, Brazil}
\acmISBN{979-8-4007-2213-4/2026/11}
\acmDOI{10.1145/3767308.3837724}

\begin{document}

\title{ReART: Reference-Guided Retrieval and Refinement for Emotion-Aware Art Generation}


\author{Qianqian Tang}
\email{qianqian.tang@whu.edu.cn}

\affiliation{%
  \institution{Wangxuan Institute of Computer\\
  Technology, Peking University}
  \city{Beijing}
  \country{China}
}

\affiliation{%
  \institution{School of Computer Science,
  Wuhan University}
  \city{Wuhan}
  \country{China}
}

\author{Jiayi Gao}
\email{gaojiayi26@stu.pku.edu.cn}

\affiliation{%
  \institution{Wangxuan Institute of Computer\\
  Technology, Peking University}
  \city{Beijing}
  \country{China}
}

\author{Ting Lei}
\email{ting\_lei@pku.edu.cn}

\affiliation{%
  \institution{Wangxuan Institute of Computer\\
  Technology, Peking University}
  \city{Beijing}
  \country{China}
}

\author{Yang Liu}
\authornote{Corresponding author.}
\email{yangliu@pku.edu.cn}

\affiliation{%
  \institution{Wangxuan Institute of Computer\\
  Technology, Peking University}
  \city{Beijing}
  \country{China}
}


\renewcommand{\shortauthors}{Qianqian Tang, Jiayi Gao, Ting Lei, and Yang Liu}

\begin{abstract}
Emotion-aware artistic image generation requires a model to satisfy semantic content, artistic style, and target emotion simultaneously. The key challenge is that artistic captions conflate these axes into underspecified free-form text, making fine-grained visual attributes such as brushwork, composition, and tonal atmosphere difficult to ground concretely. We present ReART, a reference-guided retrieval and refinement framework. Our method decomposes test captions and each image annotation in the EmoArt database into structured visual fields, and performs field-wise retrieval over subject, layout, brush-line, and tone-mood dimensions to retrieve role-specific visual references that supply the perceptual detail text alone cannot convey; these references are used alongside a structured prompt for initial synthesis. For samples where any Attribute Alignment Score (AAS) axis falls below threshold, an AAS-driven refinement loop diagnoses failures, constructs constrained repair plans specifying elements to keep, errors to fix, and operations to avoid, routes references by correction purpose, and performs controlled editing under structural preservation constraints. Our system ranks 2nd in Track 1 of the AffectiveArt 2026 Grand Challenge, achieving a perfect AAS of 1.00 and an overall score of 0.78. Code is available at https://github.com/oceanflowlab/ReART.git.
\end{abstract}



\maketitle

\section{Introduction}

Text-to-image generation has seen remarkable progress recently~\cite{wang2024review}, with diffusion-based models such as Stable Diffusion~\cite{podell2024sdxl} and DALL-E~\cite{ramesh2021zero} achieving high-fidelity synthesis across diverse styles and subjects. Among the emerging directions in this space, emotion-aware artistic image generation has attracted growing interest, as art serves not only as a means of visual representation but also as a vehicle for emotional expression~\cite{zhang2025emoart, xieaffectiveart}. In this task, given a multimodal prompt specifying semantic content, artistic style, and a target emotion, the model is expected to generate an artwork in which emotion is realized through coordinated visual decisions such as brushwork, color harmony, tonal contrast, and compositional balance rather than through explicit object depiction~\cite{yang2023emoset}.

\begin{figure*}[t]
  \centering
  \includegraphics[width=\textwidth]{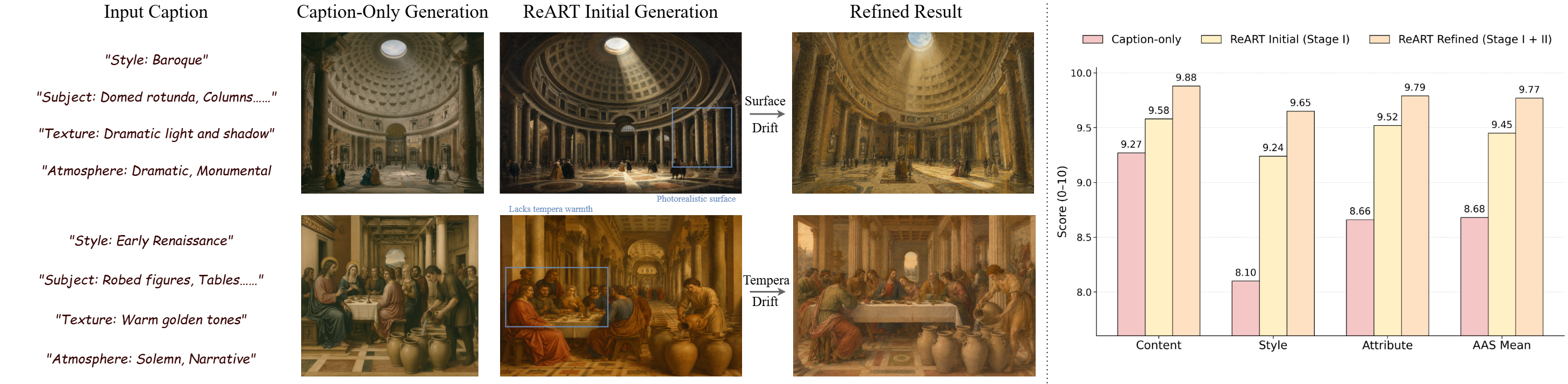}
  \caption{\textbf{Overview of ReART.} Left: qualitative examples of the proposed pipeline. For each input caption, we compare caption-only generation, ReART Stage~I generation, and the refined result after Stage~II. Highlighted regions indicate axis-specific errors that are corrected by targeted refinement while preserving already aligned content and layout. Right: AAS sub-axis scores on Local200, where ReART progressively improves alignment from caption-only generation to Stage~I and Stage~I+II.}
  \label{fig:teaser}
\end{figure*}

Achieving this, however, gives rise to two core challenges.
First, emotion is rarely explicit; it is conveyed indirectly through fine-grained visual attributes such as brushwork, color, composition, line, and texture~\cite{zhang2025emoart, yang2023emoset}. Yet captions often compress content, style, emotion, and fine-grained attributes into a single free-form sentence, leaving affective descriptions underspecified and hard to execute~\cite{peng2026survey}. For instance, as illustrated in Figure~\ref{fig:teaser}, the caption ``Style: Early Renaissance... Atmosphere: Solemn, Narrative'' specifies an emotional target but leaves unspecified how solemnity should manifest visually---whether through restrained color saturation, flattened spatial recession, or subdued surface treatment. Faced with such underspecified prompts, generative models tend to approximate a loose emotional impression while missing the fine-grained visual execution that authentic artistic affect demands.

Second, even with structured prompts and retrieved references, simultaneously satisfying content, style, emotion, and fine-grained attributes in a single pass remains difficult~\cite{yang2024emogen, dang2025emoticrafter}. As illustrated in Figure~\ref{fig:teaser}, a Baroque interior may correctly realize architectural content and spatial composition yet suffer from surface drift, rendering with a photorealistic finish rather than the painterly medium the style demands. Such failures are axis-specific; blind resampling carries no diagnostic information, while unconstrained global editing risks corrupting already-correct regions~\cite{wang2024genartist, jaiswal2026iterative}. A mechanism that pinpoints and corrects failing axes without disrupting correctly realized dimensions is therefore required.

To address these challenges, we propose \textbf{ReART}, a two-stage reference-guided framework that achieves fine-grained emotion-aware artistic generation by grounding abstract affective descriptions in structured visual evidence and correcting axis-specific failures through targeted refinement. 
 
To address the first challenge, we decompose each caption into independently addressable visual fields---subject, brushwork, color tone, composition, mood, and material surface---drawing on established art-psychological dimensions of visual emotion expression~\cite{machajdik2010affective, zhang2025emoart}
, so that each dimension can be described and matched separately. However, structured text alone is insufficient: fine-grained visual qualities such as the ink-bleed texture of a brushstroke or the tonal gradation of a specific period's palette are difficult to specify precisely in language; as prior retrieval-augmented generation work has shown, text descriptions are inherently limited in conveying rare or fine-grained visual concepts, and retrieved visual evidence is necessary to supply the perceptual detail that language leaves underspecified~\cite{chen2022re, shalev2025imagerag, yuan2025finerag}. 
We therefore retrieve field-aligned artworks from a style-specific reference pool, applying the same structured schema to reference annotations; each retrieved artwork is supplied alongside its structured field description, enabling the generative model to draw on it selectively for the dimension it best represents rather than as a holistic style template.

To address the second challenge, we introduce an Attribute Alignment Score (AAS)-driven Agentic Constrained Refinement Loop, where AAS measures alignment along content, style, and attribute axes. For samples that fail on one or more alignment axes, a diagnosis module identifies the dominant failure type; a planner constructs a structured repair plan specifying elements to keep unchanged, errors to fix, operations to avoid, and a processing priority order; a reference router selects only references relevant to the identified failure axis; and a constrained editor executes the repair under structural preservation constraints. Together, these two stages yield outputs that are close to the real artwork distribution. As shown in Figure~\ref{fig:teaser} (right), Stage~I substantially improves Style and Attribute alignment over caption-only generation by grounding abstract descriptions in field-aligned visual references, while Content is already well-handled by structured prompting alone. Stage~II further lifts all three axes, with Style showing the largest gain, demonstrating that the refinement loop effectively corrects residual axis-specific failures after initial synthesis.
 
Our method ranks 2nd overall in the AffectiveArt 2026 Grand Challenge Track~1~\cite{xieaffectiveart}, with an overall score of 0.78. Performance is evaluated via two metrics: the Fr\'{e}chet Inception Distance (FID), which quantifies distributional similarity to real artworks, and the Attribute Alignment Score (AAS), which assesses content, style, and attribute alignment. We achieve a perfect AAS of 1.00 across all sub-axes and an FID of 77.47.

 
In conclusion, we propose ReART, a two-stage Reference-guided Art generation framework with structured
Retrieval and Targeted refinement. Our contributions are threefold:

\begin{itemize}
    \item A structured visual field decomposition with field-wise
    retrieval augmented generation, grounding abstract affective captions in direct field-aligned visual evidence for synthesis.
    \item An AAS-driven diagnosis-and-repair loop with structured
    keep/fix/avoid/priority plans and purpose-routed reference selection, enabling targeted correction of axis-specific failures without disrupting correctly realized dimensions.
    \item ReART achieves a perfect AAS of 1.0 and ranks 2nd on the AffectiveArt 2026 Grand Challenge Track~1, validating the effectiveness of our framework for fine-grained emotion-aware artistic image generation.
\end{itemize}

\begin{figure*}[t]
  \centering
  \includegraphics[width=0.95\textwidth]{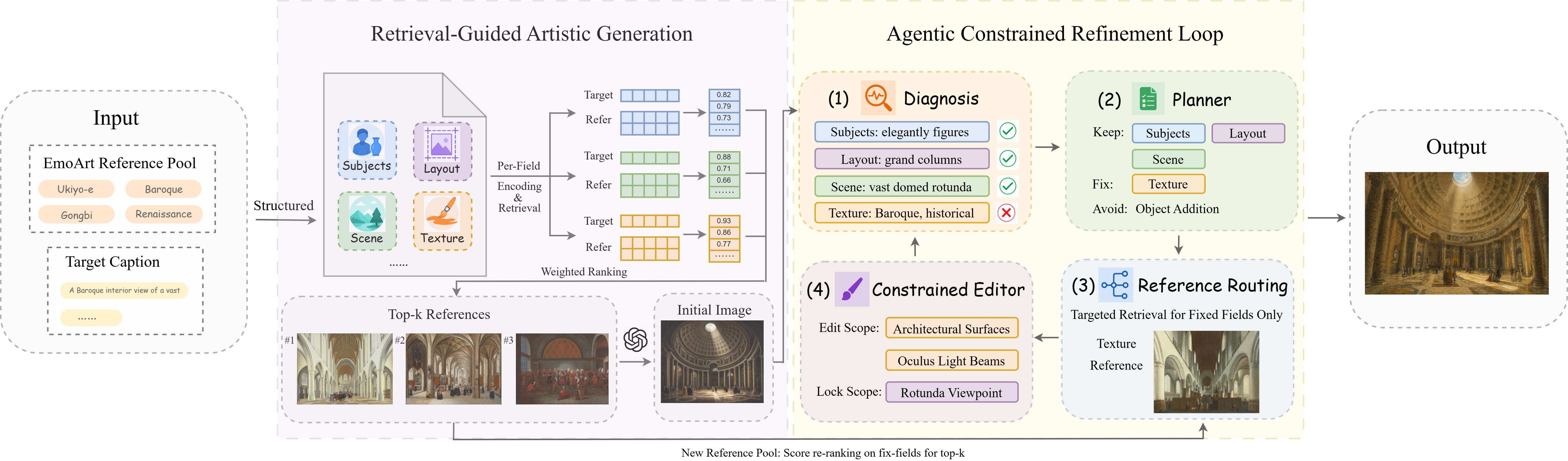}
    \caption{\textbf{Pipeline of ReART for emotion-aware artistic image generation.} ReART adopts a two-stage framework: Stage~I retrieves field-aligned references from structured visual records to guide generation, while Stage~II uses AAS-driven diagnosis and constrained editing for targeted refinement.}
  \label{fig:pipeline}
\end{figure*}

\section{Related work}

\subsection{Affective Artistic Image Generation}
Early work on visual emotion analysis focuses on recognizing affective states from images using handcrafted or learned features~\cite{yang2023emoset,wang2024expression}.
Building on large-scale diffusion models, recent generative approaches have begun to treat emotion as a controllable target: EmoGen~\cite{yang2024emogen} constructs an emotion space aligned with CLIP to generate content from discrete emotion tags, while
EmotiCrafter~\cite{dang2025emoticrafter} extends this to continuous valence-arousal control for more nuanced affective generation. EmoEdit~\cite{yang2025emoedit} further introduces an emotion adapter to manipulate existing images toward a target emotional state. However, these methods treat emotion as a global signal and largely ignore the fine-grained visual attributes---brushwork, composition, color, and tonal atmosphere---through which emotion is specifically expressed in artistic styles. ReART addresses this gap by decomposing emotional targets into independently addressable visual fields and grounding each in concrete artistic evidence.
 
\subsection{Reference-Guided Generation}


Reference-guided generation has emerged as an effective paradigm for injecting fine-grained visual concepts that text prompts alone cannot fully specify. IP-Adapter~\cite{ye2023ip} and StyleAligned~\cite{hertz2024style} transfer visual styles from reference images via cross-attention, while Re-Imagen~\cite{chen2022re} and ImageRAG~\cite{shalev2025imagerag} retrieve image-text pairs from external knowledge bases to ground rare or underspecified entities. However, these methods treat references as holistic templates and lack mechanisms to route different references to different visual dimensions. FineRAG~\cite{yuan2025finerag} improves retrieval through fine-grained image selection but still operates at the image level within a single retrieval pool. In contrast, ReART decomposes captions into visual fields and performs field-wise retrieval, independently routing references for subject, layout, brushwork, and tone-mood, with each reference serving a specific visual role.

\begin{table*}
\centering
\caption{Comparison with baseline methods on Local200. AAS is the mean of Content, Style, and Attribute alignment scores.}
\label{tab:baseline}
\begin{tabular}{lccccc}
\toprule
Method & FID $\downarrow$ & AAS (0-10) $\uparrow$ & Content $\uparrow$ & Style $\uparrow$ & Attribute $\uparrow$ \\
\midrule
FLUX.1-dev~\cite{blackforest2024flux} + IP-Adapter~\cite{ye2023ip, instantx2024fluxipadapter} & 170.51 & 6.52 & 6.78 & 7.46 & 5.33 \\
Qwen-Image~\cite{wu2025qwen} & 162.94 & 7.69 & 7.82 & 7.67 & 7.59 \\
HiDream-O1-Image~\cite{cai2026hidream} & 156.33 & 7.65 & 8.06 & 7.95 & 6.94 \\
\midrule
ReART (Ours) & \textbf{147.43} & \textbf{9.77} & \textbf{9.88} & \textbf{9.65} & \textbf{9.79} \\
\bottomrule
\end{tabular}
\end{table*}

\subsection{Agentic Image Refinement}
Recent work has explored agentic and iterative refinement as an alternative to single-pass synthesis~\cite{gao2026taming}. GenArtist~\cite{wang2024genartist} and Maestro~\cite{wan2025maestro} use multimodal LLMs to critique generated images and produce interpretable edit signals, while Agent Banana~\cite{ye2026agent} formulates editing as an agentic perception-reasoning-action loop with structured tool use. Iterative refinement methods~\cite{jaiswal2026iterative} further show that VLM-guided correction improves compositional alignment. However, these approaches mainly target user-instruction-driven editing with explicitly specified goals, whereas emotion-aware artistic generation requires automatic multi-axis alignment across content, style, and fine-grained attributes, where failures are implicit and axis-specific rather than user-identified. ReART addresses this by introducing an AAS-driven Agentic Constrained Refinement Loop that automatically diagnoses axis-specific failures, constructs structured keep/fix/avoid/priority repair plans, and routes references matched to the failing dimension for targeted correction while preserving correctly realized axes.
\section{Methodology}
\label{sec:method}

Figure~\ref{fig:pipeline} presents the overall framework of ReART, which proceeds in two stages. In Stage~I, to ground abstract affective captions in concrete visual evidence, each test caption and the EmoArt reference annotations~\cite{zhang2025emoart} are converted into structured visual records, and field-wise retrieval is performed over a style-specific reference pool; the retrieved artworks are supplied alongside their structured field descriptions and per-field similarity scores, with meta-instructions directing \texttt{gpt-image-2} to draw on each reference for the visual dimensions it best covers. 
The generated image is then evaluated across content, style, and attribute alignment using a local AAS proxy evaluator following the official challenge specification. Specifically, we instantiate the standardized LLM-assisted multimodal evaluation protocol with GPT-5.4, using fixed evaluation prompts, predefined scoring rubrics, and consistent inference settings. Images with all three sub-scores at or above 9 (on a scale of 1--10) are accepted as final outputs, while those with any sub-score below the threshold enter Stage~II, an agentic constrained refinement loop consisting of diagnosis, repair planning, reference re-routing, and controlled editing.

\subsection{Retrieval-Guided Artistic Generation}

\subsubsection{Structured Visual Record Construction}
Artistic captions typically compress multiple visual signals into a single sentence, making them difficult to disentangle for either retrieval or generation. Drawing on art-psychological dimensions of visual emotion expression~\cite{machajdik2010affective, zhang2025emoart}, we use GPT-5.4~\cite{openai2026gpt54} to decompose each test caption and each EmoArt reference annotation into a unified structured visual record, specifically decomposing these entangled signals into eleven independently addressable visual fields: main subjects, spatial relations, scene or setting, composition and layout, material or surface, brushstroke and texture, line quality, color and tone, light and shadow, and mood or atmosphere. Applying the same schema to both test captions and reference annotations places them under a unified structured representation that enables direct field-level comparison between the generation target and candidate references. 

\subsubsection{Style-Specific Pool and Field-Wise Retrieval}

To reduce retrieval noise caused by the large variation in visual attributes across artistic styles, we construct style-specific reference pools and perform field-wise retrieval within each pool. Since the fine-grained visual traits of each artistic style---such as characteristic brushwork, surface treatment, and compositional conventions---are difficult to fully capture in text descriptions, visual references drawn from a stylistically consistent pool provide essential perceptual cues that ensure stylistic consistency between retrieved references and the target artwork. 
Test samples are therefore matched to a corresponding style-specific reference pool by extracting the style keyword from the caption. 
For styles with sparse in-domain coverage, we augment the pool: the reference pool for Ink and wash is supplemented with a China-images auxiliary set, 
and the reference pool for the Gongbi style
is expanded with paintings sourced from the Chinese-Painting-Dataset~\cite{chen2017chinesepainting}, processed through the same GPT-5.4 structured compression pipeline to ensure annotation consistency. Rather than collapsing all visual fields into a single query, we encode each field independently with \texttt{BAAI/bge-m3}~\cite{chen2024bge} and compute cosine similarity between the test record and each candidate, retaining per-field scores. The final retrieval score is obtained by weighted fusion:
\begin{equation}
s(r \mid x) = \sum_{f \in \mathcal{F}} w_f \cdot s_f(r \mid x),
\end{equation}
where $s_f(r \mid x)$ is the normalized cosine similarity on field $f$ and $w_f$ is its predefined weight, with higher weights assigned to semantically critical fields such as main subjects and composition, and lower weights to auxiliary cues such as line quality and mood. 

For samples with strong layout cues such as hanging scroll or fan-shaped formats, we apply a layout-heavy weight scheme that increases the weights on composition and scene fields while proportionally reducing weights on less critical fields. Candidates are ranked by the fused score and the top-10 references are retained for each test sample.


\subsubsection{Reference-Augmented Generation}

To ground abstract affective requirements in concrete visual evidence, we construct a structured generation prompt for \texttt{gpt-image-2}~\cite{openai2026gptimage2}, which accepts both text and reference images as multimodal inputs. Given the structured test record and top-10 retrieved references, the prompt consists of five blocks: \textbf{(1) Task}, which preserves the original caption as the primary objective; \textbf{(2) Visual Attributes}, which specifies the target visual fields; \textbf{(3) Structured Test Record}, which provides the structured representation; \textbf{(4) Reference Usage}, which associates each reference with its field-level descriptions and guides selective usage for supported dimensions such as composition, brush texture, and color atmosphere; and \textbf{(5) Negative Constraints}, which prevent content, format, and composition drift as well as over-polished digital rendering. This design enables references to provide role-specific visual evidence while keeping the caption as the primary source of truth.

\subsection{Agentic Constrained Refinement Loop}

\begin{table*}
\centering
\caption{Ablation study on Local200 validating the contribution of each pipeline component across FID and AAS sub-axis scores.}
\label{tab:ablation}
\begin{tabular}{lccccc}
\toprule
Method & FID $\downarrow$ & AAS (0-10) $\uparrow$ & Content $\uparrow$ & Style $\uparrow$ & Attribute $\uparrow$ \\
\midrule
(1) Caption-only generation                  & 193.47 & 8.68 & 9.27 & 8.10 & 8.66 \\
(2) Structured fields, no references         & 186.86 & 8.84 & 9.36 & 8.38 & 8.77 \\
(3) Structured fields + references (Stage I) & 158.69 & 9.45 & 9.58 & 9.24 & 9.52 \\
(4) Naive Re-edit                             & 157.73 & 9.45 & 9.62 & 9.17 & 9.54 \\
(5) w/o Constrained Editing                   & 154.63 & 9.60 & 9.73 & 9.40 & 9.67 \\
(6) w/o Reference Routing                     & 150.78 & 9.67 & 9.83 & 9.47 & 9.71 \\
(7) Full pipeline (Stage I + II)              & \textbf{147.43} & \textbf{9.77} & \textbf{9.88} & \textbf{9.65} & \textbf{9.79} \\
\bottomrule
\end{tabular}
\end{table*}

\subsubsection{AAS Scoring and Sample Selection}

To identify samples that require targeted refinement, we evaluate each generated image 
with a local AAS evaluator implemented with GPT-5.4 following the official evaluation protocol.
In the AffectiveArt challenge, AAS is defined as a standardized LLM-assisted multimodal evaluation protocol over three aspects: content alignment, style alignment, and attribute alignment. Because the organizer-side evaluation backend is not publicly released, we approximate it with GPT-5.4 under fixed evaluation prompts, predefined scoring rubrics, and consistent inference settings. Content alignment focuses on whether required subjects, scene type, and spatial relations are correctly depicted, corresponding to the main subjects, scene, and spatial-relation fields in our structured record. Style alignment measures whether the intended artistic style, medium logic, material surface, and period-consistent visual language are preserved. Attribute alignment evaluates whether fine-grained visual attributes, including composition, brushwork, line quality, color tone, lighting, and mood atmosphere, are properly realized. Any image with at least one sub-score below 9 (on a scale of 1--10) is selected for the refinement loop; the rest are accepted as final outputs.

\subsubsection{Diagnosis}

To pinpoint axis-specific failures and define what should be preserved during editing, we propose a diagnosis module that identifies dominant error types and repair boundaries. The diagnosis module receives the current generated image, the original caption, the structured visual record, and the per-field retrieval scores from Stage~I, all provided as context in a single multimodal prompt to GPT-5.4. It determines which structured field constraints are already satisfied and should be preserved, which constraints have failed, what the dominant error type is among subject error, layout error, texture error, tone error, and mixed error, and what the appropriate repair scope and protected scope are. The per-field retrieval scores are included to indicate which artworks in the initial top-10 are most relevant to each visual dimension, and are later used by the router to select references for the failed dimensions.

\subsubsection{Planner}

To transform unconstrained resampling into an interpretable repair procedure, GPT-5.4 constructs a repair plan based on the diagnosis output. The plan consists of four components: (1) \textbf{Keep}: preserves correctly realized elements during editing, including subject identity, spatial relations, composition skeleton, and page format. (2) \textbf{Fix}: specifies field-level errors identified by diagnosis, such as incorrect texture, missing medium characteristics, inaccurate color temperature, or weak line quality, together with their target repair scopes. (3) \textbf{Avoid}: defines forbidden operations that may damage correct content or structure, such as adding unsupported objects, changing scene types, drifting from page format, or breaking composition. (4)  \textbf{Priority}: orders repairs from structural to aesthetic refinement, prioritizing subject identity and spatial relations, followed by layout and appearance refinements.

\subsubsection{Reference Routing and Constrained Editing}

To supply axis-specific visual evidence for each repair target without introducing irrelevant style interference, we propose a reference routing mechanism that selects references from the initial top-10 based on field-level scores. Rather than reusing all retrieved references uniformly, for each dimension identified in \textbf{Fix}, we select the four candidates with the highest field-level scores on that dimension as the routed references for the corresponding repair axis. The constrained editor then performs targeted image editing with \texttt{gpt-image-2}, taking the current image as the primary canvas and the routed references as axis-specific visual evidence. We implement the lock scope through explicit textual edit constraints, instructing the model to preserve the subject core, silhouette, composition skeleton, and major spatial relations while modifying only the failed visual dimension. The refined image is re-evaluated by the AAS evaluator, and the loop repeats until all three sub-scores reach the threshold or the maximum iteration budget is reached.

\section{Experiment}

\subsection{Experimental Setup}

\paragraph{Datasets and Evaluation Protocol.}

We conduct experiments under two evaluation settings. For the \textbf{official evaluation}, the AffectiveArt 2026 Grand Challenge Track~1 provides 1,000 test captions and evaluates submissions against a hidden reference set of 2,000 artworks constructed by the organizers to assess affective generation capability. The overall score is computed as $0.5 \times \text{FID Score} + 0.5 \times \text{AAS}$, where FID~\cite{heusel2017gans} is first normalized before combination. Because the official evaluation backend and hidden reference set are not accessible, it is not possible to perform controlled ablation or baseline comparison under the official protocol. We therefore construct a \textbf{Local200} benchmark by sampling 200 images from the EmoArt reference pool to cover the artistic styles present in the challenge, serving as a local validation set for controlled experiments; ground-truth artworks are retained for FID~\cite{heusel2017gans} computation. AAS on Local200 is approximated with our local GPT-5.4 evaluator under fixed prompts, predefined rubrics, and consistent inference settings, as described in Section~\ref{sec:method}. Local200 is used solely for controlled ablation and baseline comparisons; results are not directly comparable to the official leaderboard scores.

\paragraph{Implementation Details.}
All methods use the same preprocessing pipeline, structured field text, and retrieved references. For field-wise retrieval, each visual field is encoded with \texttt{BAAI/bge-m3}~\cite{chen2024bge}, and candidates are ranked by the weighted fusion of normalized cosine similarities, with higher weights assigned to semantically critical fields such as main subjects and composition. The top-10 references are retained for each test sample. For ReART, both initial synthesis and constrained editing are performed with \texttt{gpt-image-2}~\cite{openai2026gptimage2}. The refinement loop uses an AAS threshold of 9 for each sub-axis and runs for at most four iterations. FLUX.1-dev~\cite{blackforest2024flux} + IP-Adapter~\cite{ye2023ip, instantx2024fluxipadapter}, Qwen-Image~\cite{wu2025qwen}, and HiDream-O1-Image~\cite{cai2026hidream} are evaluated with the same structured field text and retrieved references under their corresponding inference settings.

\subsection{Comparison with Baselines}
\begin{figure}[h]
\centering
\includegraphics[width=0.95\columnwidth]{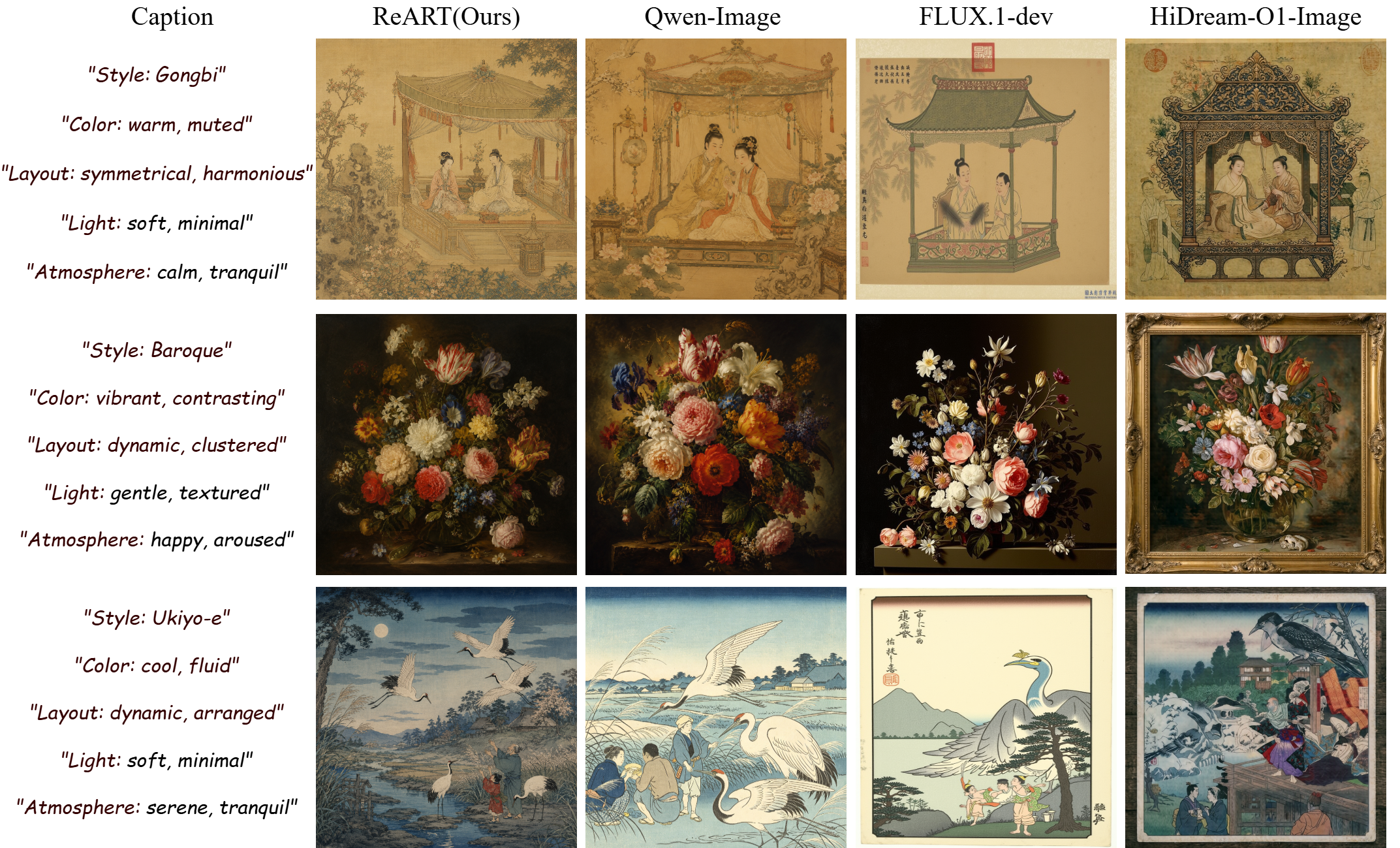}
\caption{Qualitative comparisons with baseline methods on three representative samples.}
\label{fig:qualitative}
\end{figure}

\begin{table*}
\centering
\caption{Top-5 results on the official AffectiveArt 2026 Grand Challenge Track~1 leaderboard. EmoForge is our submitted system.}
\label{tab:leaderboard}
\begin{tabular}{clcccccc}
\toprule
Rank & Team & Overall $\uparrow$ & FID $\downarrow$ & AAS (0-1) $\uparrow$ & Content $\uparrow$ & Style $\uparrow$ & Attribute $\uparrow$ \\
\midrule
1 & USTC\_PI\_LAB\_TEAM  & \textbf{0.80} & \textbf{66.12} & 0.99 & 0.99 & 0.99 & 1.00 \\
2 & EmoForge (Ours)      & 0.78 & 77.47 & \textbf{1.00} & \textbf{1.00} & \textbf{1.00} & \textbf{1.00} \\
3 & VIRlab               & 0.78 & 75.56 & 0.99 & 0.98 & 0.99 & 0.99 \\
4 & edaich               & 0.77 & 78.59 & 0.99 & 0.98 & 0.99 & 0.99 \\
5 & Latent Feeling        & 0.76 & 87.51 & 1.00 & 0.99 & 1.00 & 1.00 \\
\bottomrule
\end{tabular}
\end{table*}
Table~\ref{tab:baseline} and Figure~\ref{fig:qualitative} present the quantitative and qualitative results on Local200, respectively. ReART achieves the best performance across all metrics, obtaining an FID~\cite{heusel2017gans} of 147.43, which is 8.9 points lower than that of the strongest baseline, HiDream-O1-Image~\cite{cai2026hidream}, and an AAS of 9.77, representing a 27.0\% relative improvement over Qwen-Image~\cite{wu2025qwen} (7.69). Compared with the best-performing baseline on each sub-axis, ReART improves Content, Style, and Attribute from 8.06, 7.95, and 7.59 to 9.88, 9.65, and 9.79, respectively, demonstrating its effectiveness in improving both distributional similarity and alignment across content, style, and fine-grained attributes.

A closer comparison shows that FLUX.1-dev~\cite{blackforest2024flux} + IP-Adapter~\cite{ye2023ip,instantx2024fluxipadapter} still yields the highest, and thus worst, FID (170.51) despite using the same reference images, indicating that holistic reference conditioning may introduce visual cues inconsistent with the target style. In contrast, ReART uses field-wise routing so that different references contribute to the visual dimensions they best support. The qualitative results further confirm this advantage: for the Gongbi sample, ReART better preserves the symmetrical composition, warm muted tones, and delicate brushwork; for the Baroque floral sample, it more accurately captures the vivid color contrasts and rich painterly texture; and for the Ukiyo-e landscape sample, it better reproduces the cool, fluid tones and serene atmosphere. In comparison, the baseline methods are more prone to style drift, over-smoothed rendering, and the loss of medium-specific surface characteristics.

\subsection{Ablation Study}

Table~\ref{tab:ablation} reports the ablation results on Local200, isolating the contribution of each component in the ReART pipeline.

\textbf{Effect of structured field decomposition.} Moving from caption-only generation (1) to structured fields without references (2) yields a consistent improvement: FID~\cite{heusel2017gans} drops from 193.47 to 186.86, while AAS increases from 8.68 to 8.84. The gains concentrate more on Style and Attribute than on Content, indicating that field decomposition improves fine-grained visual control even without external visual evidence.

\textbf{Effect of field-wise reference retrieval.} Adding field-aligned reference images (3) produces the largest single improvement: FID drops from 186.86 to 158.69, and AAS rises from 8.84 to 9.45. The improvements are especially pronounced on Style and Attribute, demonstrating that retrieved references provide perceptual details that structured text descriptions alone cannot convey.

\textbf{Effect of AAS-driven refinement.} The full pipeline (7) further reduces FID from 158.69 to 147.43 and improves AAS from 9.45 to 9.77. Among the 200 Local200 samples, 42 images entered refinement after Stage~I, requiring an average of 1.38 iterations, with 71.4\% converging within one iteration. The Style axis was the most frequent failure dimension, where images were semantically correct but suffered from over-polished digital rendering. The refinement loop corrects such failures through targeted reference routing and constrained editing, improving Style alignment from 9.24 to 9.65.

To further analyze Stage~II, we compare several variants. Naive Re-edit (4), which performs an unconstrained second editing pass without diagnosis, planning, routing, or preservation constraints, achieves only 9.45 AAS, showing that simple re-editing is insufficient. Removing constrained editing (5) or reference routing (6) decreases AAS to 9.60 and 9.67, respectively, confirming that structural preservation and purpose-specific reference selection are both important for effective refinement.

\subsection{Official Challenge Results}

Table~\ref{tab:leaderboard} reports the top-5 results on the official AffectiveArt 2026 Grand Challenge Track~1 leaderboard. ReART (submitted as EmoForge) ranks 2nd overall with an overall score of 0.78, achieving a perfect AAS of 1.00 across all three sub-axes---the highest AAS among all submitted systems. The 1st-place team USTC\_PI\_LAB\_TE\\AM achieves a higher overall score of 0.80, driven by a lower FID of 66.12 compared to our 77.47, while our AAS of 1.00 surpasses their 0.99. These results confirm that the structured retrieval and targeted refinement approach of ReART excels at fine-grained prompt-image alignment, while further improvement in distributional similarity to real artworks remains an avenue for future work.

\section{Conclusion}

In this paper, we present ReART, a two-stage reference-guided framework for emotion-aware artistic image generation. To address the semantic ambiguity of artistic captions, we decompose captions into structured visual records and perform field-wise retrieval over EmoArt reference pools, enabling retrieved artworks to serve as role-specific visual evidence rather than holistic style templates. To correct axis-specific alignment failures after synthesis, we introduce an AAS-driven agentic refinement loop that diagnoses failures, constructs keep/fix/avoid/priority repair plans, routes references matched to failing dimensions, and performs constrained editing under structural preservation constraints. On the AffectiveArt 2026 Grand Challenge Track~1, ReART achieves a perfect AAS of 1.00 on all three sub-axes and ranks 2nd overall, demonstrating the effectiveness of structured retrieval and targeted refinement.

\begin{acks}
This work was supported by the grants from the National Natural Science Foundation of China 62372014 and the Beijing Nova Program.
\end{acks}


\bibliographystyle{ACM-Reference-Format}
\bibliography{reference}
\appendix

\end{document}